\documentclass[aps, prb, twocolumn, superscriptaddress, floatfix]{revtex4-2} 
\usepackage{etex}

\usepackage{amsmath, amssymb, amsfonts, braket, commath, bm, empheq, array, lipsum}

\usepackage[]{graphicx}
\DeclareGraphicsExtensions{.eps,.jpg,.png,.pdf}
\usepackage{hyperref, url}
\hypersetup{
	colorlinks=true,
	linkcolor=red,
	filecolor=blue,
	urlcolor=magenta,
}

\usepackage{dblfloatfix}

\newcommand{\Capheat}{\ensuremath{\mathcal{C}}}

\newcommand\Eq{\ensuremath{E_q}}

\newcommand\FC{ \ensuremath{F} }

\newcommand\iiserk{Department of Physical Sciences, Indian Institute of Science Education and Research Kolkata, Mohanpur 741246, India}

\newcommand\lboro{School of science, Loughborough University, Loughborough, Leicestershire LE11 3TU, United Kingdom}

\newcommand\mean[1]{\ensuremath{\left\langle #1 \right\rangle}}

\newcommand{\nrm}{\ensuremath{Z}}

\newcommand{\prob}[1]{\ensuremath{\textrm{P}\del{#1}}}
\newcommand\phaseA{annealed}
\newcommand\phaseQ{quenched}
\newcommand\phaseF{frozen}

\newcommand\SMC{ \ensuremath{\mathcal{S}} }

\newcommand{\sff}[1]{\ensuremath{\mathcal{K}\del{#1}}}
\newcommand{\sffdc}[1]{\ensuremath{\mathcal{K}_\mathrm{dc}\del{#1}}}
\newcommand{\sffc}[1]{\ensuremath{\mathcal{K}_\mathrm{c}\del{#1}}}
\newcommand\sffbar{ \ensuremath{ \overline{\mathcal{K}} } }
\newcommand{\sgE}{\ensuremath{\sigma_{E}}}

\newcommand\tdip{\ensuremath{t_\mathrm{dip}}}

\newcommand\Tc{\ensuremath{T_c}}
\newcommand\Tq{\ensuremath{T_q}}
\newcommand\TMBL{\ensuremath{T_\mathrm{MBL}}}

\begin{document}
\title{Identifying mobility edge from finite temperature spectral form factor}

\author{Basudha Roy}\email{br23rs008@iiserkol.ac.in}
\affiliation{\iiserk}
\author{Adway Kumar Das}\email{A.K.Das@lboro.ac.uk}
\affiliation{\lboro}
\author{Anandamohan Ghosh}\email{anandamohan@iiserkol.ac.in}
\affiliation{\iiserk}

\begin{abstract}
The spectral form factor (SFF) is a measure of energy correlations and has been widely used to identify the transition from the ergodic to the localized phase in interacting many-body quantum systems. In this work, we show that in a disordered Heisenberg spin-$\frac{1}{2}$ model, the finite temperature SFF can be used to generate a canonical phase diagram exhibiting a critical temperature $\TMBL$. Using simple ideas of statistical mechanics, we obtain the critical energy density $\epsilon_\mathrm{MBL}$ dual to $\TMBL$. We show that the mobility edge, numerically estimated from the spread of local perturbations and the optical conductivity, indeed coincides with  $\epsilon_\mathrm{MBL}$.
\end{abstract}
\maketitle

\section{Introduction}
In interacting many-body quantum systems, disorder-induced localization is a well-established paradigm to break ergodicity~\cite{Basko2006, Pal2010, Nandkishore2015}. In such a many-body localized (MBL) phase, particle or energy transport is absent due to conserved quasilocal charges~\cite{Serbyn2013, Abanin2019, Chertkov2021}. The interplay between the interaction and disorder drives a dynamical phase transition from the MBL to an ergodic phase. The analogous phenomenon in the single-particle systems is the Anderson transition~\cite{Anderson1958, Evers2008, Lahini2008, Nosov2019, Sierant2020, Das2022, Das2022b, Das2024}. Importantly, at weak disorder strength, the excited ergodic states remain separated from the localized states by a critical energy known as the mobility edge (ME)~\cite{Mott1967, AbouChacra1973, AbouChacra1974, Izrailev1999, Aizenman2011, Semeghini2015, Sarkar2021, Das2023, Seth2026Arxiv, Pawlik2024}.

The apparent similarity between the MBL and Anderson localization questions the existence of a ME in interacting many-body systems. Numerical studies based on the energy correlations, conductivity, and entanglement entropy demonstrate the existence of a ME on the ergodic side of the MBL transition in various many-body systems~\cite{Laumann2014, Luitz2015, Serbyn2015, Modak2015, Kohlert2019, Brighi2020, Roy2020, Das2026Arxiv}. However, avalanche instability from rare thermal bubbles across the spectrum may destabilize the ME for large system sizes inaccessible to exact diagonalization~\cite{DeRoeck2016, Potirniche2019, Suntajs2020, Gopalakrishnan2019, Dumitrescu2019, KieferEmmanouilidis2021}. Thus, it is important to construct probes of the ME which can be computed without diagonalizing the governing Hamiltonian.

In this work, we show that the ME can be identified from the finite temperature spectral form factor (SFF). Such a quantity can be calculated without doing exact diagonalization, e.g.,~via the finite temperature Lanczos method~\cite{Jaklic1994}. The SFF is experimentally accessible via non demolition measurement of ancilla qubits~\cite{Vasilyev2020}, randomized measurements~\cite{Joshi2022, Dong2025}, and molecular spectroscopy~\cite{Leviandier1986, Guhr1990, Michaille1999, Das2025}. At infinite temperature, the SFF behaves as a global observable and exhibits a correlation hole in the case of ergodic systems. By monitoring the relative depth of the correlation hole as a function of temperature, we identify a critical temperature $\TMBL$ which is dual to the ME. We verify our protocol against the spread of local perturbation~\cite{Serbyn2015} and optical conductivity~\cite{Steinigeweg2016}.

The paper is organized as follows. In Sec.~\ref{sec_canonical}, we introduce the Heisenberg model and corresponding canonical phase diagram identifying the critical temperatures separating three distinct phases. In Sec.~\ref{sec_SFF}, we discuss how to identify the mobility edge from the finite temperature SFF and corroborate our estimates from the existing measures. Our concluding remarks are given in Sec.~\ref{sec_Discussion}.

\section{Canonical phase diagram}\label{sec_canonical}
\begin{figure}[t]
	\centering
	\includegraphics[width=\columnwidth]{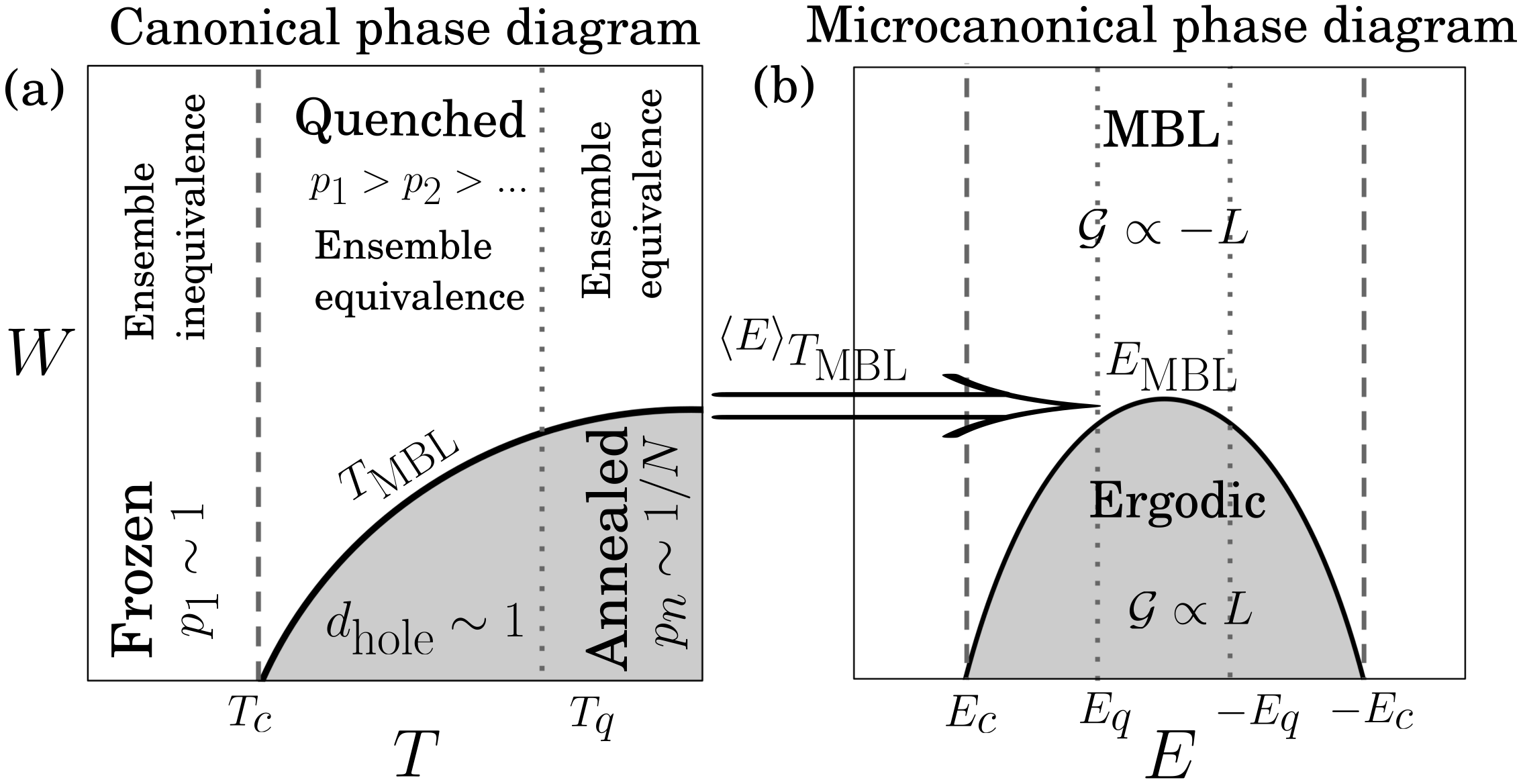}
	\caption{Canonical and microcanonical phase diagrams of a generic many-body disordered system where $W$, $T$ and $E$ are disorder strength, temperature and energy, respectively. 
    (a)~Gray dashed vertical lines indicate critical temperatures $\Tc$ and $\Tq$. The black solid line denotes the critical temperature $T_\mathrm{MBL}$, dual to ME. $d_\mathrm{hole}$ is the relative depth of the correlation hole of the finite temperature SFF. $p_n$ is the Gibbs weight of the $n$th energy state at temperature $T$.
        (b)~$\mathcal{G}$ is the Thouless conductance like quantity obtained from the spread of a local perturbation [Eq.~\eqref{eq:G_Abanin}] and $L$ is the linear size of the system. Gray dashed vertical lines indicate $E_c$ and $E_q$, dual to $\Tc$ and $\Tq$, respectively. The black solid line denotes the ME.}
	\label{fig:1}
\end{figure}

Consider the one-dimensional (1D) disordered Heisenberg spin-$\frac{1}{2}$ model with isotropic nearest-neighbor 
interaction among $L$ spins and open boundary condition~\cite{Pal2010, Luitz2015, Das2025}
\begin{align}
	\label{eq:XXX_H}
	\hat{H} &= \frac{1}{2}\sum_{j = 1}^{L} h_j \hat{\sigma}_j^z + \frac{1}{4} \sum_{k = 1}^{L-1} \vec{\sigma}_k\cdot \vec{\sigma}_{k+1}
\end{align}
where $\vec{\sigma}_j\equiv \{ \hat{\sigma}^x_j, \hat{\sigma}^y_j, \hat{\sigma}^z_j \}$ are the Pauli operators on the $j$th site and the random Zeeman splitting, $h_j$ is uniformly distributed within $[-W, W]$. The U(1) symmetry of $\hat{H}$ led us to look at the zero magnetization sector with Hilbert space dimension $N = \binom{L}{L/2}$. The Heisenberg model is a prototypical interacting many-body system for studying many-body localization~\cite{Avishai2002, Pal2010, Luitz2015, Chanda2020, Sierant2019, TorresHerrera2015, Das2025, VallejoFabila2024, VallejoFabila2025}. A corresponding microcanonical phase diagram is schematically shown in Fig.~\ref{fig:1}(b), where the ME separates the localized and extended states in the energy space.

The presence of quenched disorder in the Heisenberg model motivates us to explore another promising avenue that breaks ergodicity, namely, the glassy dynamics where the free energy landscape is rugged with many metastable configurations~\cite{Laumann2014, Garrahan2018}. A mean-field model of such nonergodic systems is the quantum random energy model (QREM)~\cite{Laumann2014, Baldwin2016, Derrida1980}. A corresponding microcanonical treatment reveals a critical energy, $E_c$, below which the replica symmetry breaks down and the system enters a \phaseF~phase where the ground state alone dictates the equilibrium properties~\cite{Derrida1981, Goldschmidt1990, Joerg2008, Das2024}. The free energy of QREM is nonanalytic at $E_c$, indicating an equilibrium phase transition to a paramagnetic phase above $E_c$. However, the ME may exist above the critical energy $E_c$ such that local order parameters remain nonzero even above $E_c$~\cite{Laumann2014}. Similar critical energy exists in the case of the Heisenberg model, as we show below.

In order to obtain a microcanonical description of the Heisenberg model, we look at the density of states (DOS)
\begin{align}
	\rho(E) = \frac{1}{N} \sum_{n = 1}^{N} \delta\del{E - E_n}.
	\label{eq:DOS_def}
\end{align}
In generic disordered systems, the ensemble-averaged DOS, $\mean{\rho(E)}$ is a smooth concave function of the energy~\cite{Livan2018book, Nakano2018, Bertuola2005, Das2025a} with exceptions being fragmented spectrum~\cite{Russomanno2021, Pal2025Arxiv, Das2022a}. The DOS of the Heisenberg model is approximately Gaussian with mean $-\frac{1}{4}$ and second moment $\mean{E^2} = \frac{3}{16}L + \frac{W^2 }{12} L$ (see Appendix~\ref{sec_appnd_E_moments}). Thus, we shift and scale the spectrum such that the DOS has zero mean and variance of $\frac{1}{4}$ irrespective of the system size and disorder strength.

Given the DOS, $\rho(E)$, and Hilbert space dimension $N$, the number of eigenstates within the energy window $(E - \frac{dE}{2}, E + \frac{dE}{2})$ is $\mathcal{N}(E) = N \rho(E) dE$ where $dE \ll 1$ is an infinitesimal width. If the DOS is self-averaging, we can ignore the sample-to-sample fluctuations and $\mean{\ln \mathcal{N}(E)} = \ln \mean{\mathcal{N}(E)}$ in the limit $N \to \infty$. Since we consider a quantum system isolated from the environment, assuming all the eigenstates within $(E - \frac{dE}{2}, E + \frac{dE}{2})$ are equally probable, the ensemble-averaged microcanonical entropy at energy $E$ is
\begin{align}
	\SMC(E) & = \begin{cases}
		0, & |E| \geq E_c\\
		\ln N + \ln \mean{\rho(E)}, & |E| < E_c
	\end{cases}
	\label{eq:S_Boltz}
\end{align}
where $\mean{\rho(E_c)} \equiv N^{-1}$, the Boltzmann constant is taken to be unity and we ignore the constant part $\ln dE$ coming from coarse-graining as $dE \ll 1 \ll N$. The non-negativity of the ensemble-averaged microcanonical entropy is responsible for its piecewise behavior and leads to nonanalyticity in the free energy, as we discuss next.

\begin{figure}[t]
	\centering
	\includegraphics[width=\columnwidth]{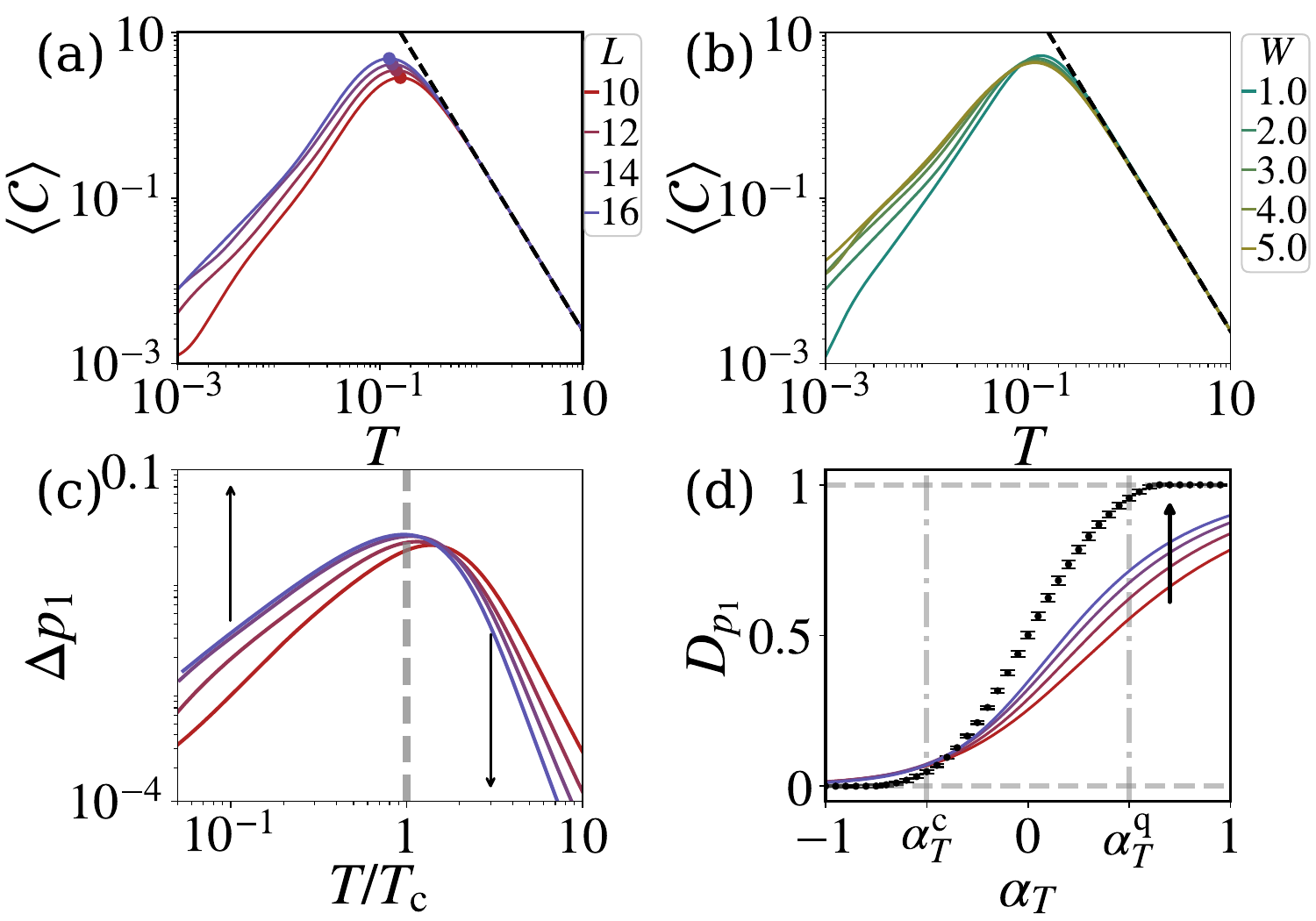}
	\caption{(a)~Ensemble-averaged heat capacity for $W = 2$ and various values of $L$. Black dashed line denotes the high-temperature behavior. 
        (b)~Same quantity as in (a) for $L = 16$ and different disorder strengths.
		(c)~Difference between the average and typical Gibbs weight of the ground state, $\Delta p_1 \equiv \left|\langle p_1 \rangle-p_1^\mathrm{typ}\right|$.
		(d)~Scaling exponent $D_{p_1}^{(N)}$ as a function of the temperature parameter $\alpha_T$ [Eq.~\eqref{eq:aT_def}] for different system sizes. Black markers correspond to $D_{p_1}$, the exponent in the thermodynamic limit where error-bars denote 95\% confidence interval. Vertical dot-dashed lines mark the two critical values of $\alpha_T$.}
	\label{fig:2}
\end{figure}

The energy is thermodynamically conjugate to the inverse temperature. Thus, Legendre transformation of the microcanonical entropy gives the canonical free energy where the slope of $\SMC(E)$ at a given energy provides the dual temperature~\cite{Ruelle1969book, kardar2007}. The free energy can also be obtained as $\FC(T) = - T \ln \nrm(T)$ where
\begin{align}
	\nrm(T) = \sum_{n = 1}^{N} e^{-\frac{E_n}{T}}
	\label{eq:Z_def}
\end{align}
is the partition function and $p_n \equiv e^{-\frac{E_n}{T}}/\nrm(T)$ is the Gibbs weight of the $n$th energy state at temperature $T$.

To understand the equilibrium phase transition of the Heisenberg model, we need to look for nonanalyticity in the free energy.  The derivatives of the free energy are functions of the thermal moments of energy, $\mean{E^k}_T \equiv \sum_n E_n^k p_n$. Particularly, the second derivative is related to the heat capacity
\begin{align}
	\Capheat \equiv -T\dpd[2]{\FC}{T} = \frac{\mean{E^2}_T - \mean{E}_T^2}{T^2}
	\label{eq:Cv_def}
\end{align}
Using annealed approximation and the fact that the DOS is approximately Gaussian, we get the ensemble-averaged thermal moments of energy as
\begin{align}
	\begin{split}
		\mean{\mean{E^k}_T} &\approx \dfrac{N\int dE \mean{\rho(E)} E^k e^{-\frac{E}{T}}} {\mean{\nrm(T)}} \\
        &= \begin{cases}
			-\frac{\sgE^2}{T}, & k = 1\\
			\sgE^2 + \frac{\sgE^4}{T^2}, & k = 2
		\end{cases}\\
	\end{split}
	\label{eq:heat_cap_annealed}
\end{align}
such that the average heat capacity at high temperature is $\mean{\Capheat} = \frac{\sgE^2}{T^2}$ where $\sgE$ is the spectral width. Contrarily, at very low temperature, the ground state with energy $E_1$ is the only allowed configuration of the spin chain such that the Gibbs weights of all the excited states are negligible compared to the ground state. Then, the partition function $\nrm(T) \sim e^{-\frac{E_1}{T}}$ and the mean free energy coincides with the ensemble average of the ground state energy, which exponentially decays with system size as $\mean{E_1} \approx -a(L) e^{-b(L) W} - c(L)$ where $W$ is the disorder strength and $a, b, c$ are fitting parameters given in Eq.~\eqref{eq:param_vals}. 

In Fig.~\ref{fig:2}(a), we show the ensemble-averaged heat capacity, $\mean{\Capheat}$, for a fixed disorder strength and different system sizes. As we shift and scale the energy spectrum such that the spectral width is independent of system size and disorder strength, $\mean{\Capheat}$ decays quadratically at high temperature independent of $W$ [Eq.~\eqref{eq:heat_cap_annealed}] as shown in Fig.~\ref{fig:2}(b). Importantly, we find that the heat capacity exhibits a maximum at a critical temperature $\Tc$ scaling as
\begin{align}
	\Tc \propto 1/\sqrt{\ln N}.
	\label{eq:Tc_scaling}
\end{align}
The critical temperatures are denoted via circular markers in Fig.~\ref{fig:2}(a). Then, Ehrenfest classification of criticality dictates that the Heisenberg model undergoes a second order phase transition at $\Tc$. The critical temperature $\Tc$ is independent of the disorder strength as shown in Fig.~\ref{fig:2}(b). We also find that
\begin{align}
    \mean{E}_{T = \Tc} \approx E_c
\end{align}
where $E_c$ is the critical energy obtained from the microcanonical treatment in Eq.~\eqref{eq:S_Boltz}. Thus, similar in spirit to the QREM, we identify the equilibrium phase transition of the Heisenberg model at $T = \Tc$, below which the \phaseF~phase exists with lack of ensemble equivalence, i.e.,~the difference between the canonical and microcanonical descriptions does not vanish in the limit $N\to\infty$~\cite{Ruelle1969book, Touchette2004, Campa2025}. The ensemble inequivalance below $\Tc$ can be verified from the difference between the typical and average behavior of the Gibbs weight of the ground state energy as shown in Fig.~\ref{fig:2}(c).


At high temperature, the partition function $\nrm(T) = N$, i.e.~all the energy states are equally probable irrespective of the nature of correlation among the energy levels. The critical temperature above which the equipartition holds is denoted by $\Tq$ such that
\begin{align}
	T > \Tq \Leftrightarrow \nrm(T) = N,\: p_n = N^{-1}\;\forall\; n 
\end{align}
The region $T > \Tq$ is denoted as the \phaseA~phase where equipartition as well as ensemble equivalence holds. The equipartition implies that the expectation value of any local observable with respect to initial states having energy within $[-\Eq, \Eq]$ follows the infinite temperature average in the ergodic phase of the Heisenberg model, where $E_q \equiv \mean{E}_{T = T_q}$. To identify the temperature $\Tq$ beyond which $\mean{p_1} = N^{-1}$, we define the following finite size scaling exponent
\begin{align}
    D_{p_1}^{(N)} \equiv -\frac{\ln \mean{p_1^{(N)}}}{\ln N}
    \label{eq:D_p1_def}
\end{align}
where $p_1^{(N)}$ is the Gibbs weight of the ground state for system size $N$ such that $\lim\limits_{N\to\infty} D_{p_1}^{(N)} \equiv D_{p_1}$. Further the system size scaling of the critical temperature [Eq.~\eqref{eq:Tc_scaling}] prompts us to scale the temperature as
\begin{align}
    T \equiv (\ln N)^{\alpha_T}.
    \label{eq:aT_def}
\end{align}
We observe that $D_{p_1}^{(N)} \approx D_{p_1} + A/\ln N$ where $A$ is a constant. Thus, we extrapolate $D_{p_1}^{(N)}$ with respect to $1/\ln N$ and find that in the limit $N \rightarrow \infty$, 
\begin{align}
	D_{p_1} = \begin{cases}
	    0, & \alpha_T < -\frac{1}{2}\\
        1, & \alpha_T \geq \frac{1}{2}.
	\end{cases} 
	\label{eq:Tq_scaling}
\end{align}
Therefore, in terms of the new parametrization of the temperature, the critical temperature exponents in the canonical phase diagram are
\begin{align}
	\alpha_T^c = -\frac{1}{2}, \quad \alpha_T^q = \frac{1}{2}.
\end{align}
In Fig.~\ref{fig:2}(d) we show the variation of $D_{p_1}$ as a function of $\alpha_T$ and clearly identify the two critical points. 
Thus, we have identified two critical temperatures, and in between the \phaseF~and \phaseA~phase there exists the \phaseQ~phase ($\Tc < T < \Tq$) where equipartition is absent but ensemble equivalence is present. In the next section, we show the dynamical manifestations of the above three thermodynamic phases.

\section{Finite temperature spectral form factor}\label{sec_SFF}

The thermodynamics discussed so far is dependent on the partition function, which cannot detect the dynamical phase transition across the mobility edge. However, if we analytically continue the partition function to the complex plane, $\nrm(T, it) = \sum_{n = 1}^{N} e^{-\del{\frac{1}{T} + it} E_n}$ where the imaginary part plays the role of time, we can identify the mobility edge. In particular, the squared modulus of the complex partition function gives the finite temperature SFF~\cite{Cotler2017a, Xu2019, DelCampo2020}
\begin{align}
	\sff{t; T} &\equiv \frac{\abs{\nrm(T,it)}^2}{\nrm(T)^2}.
	\label{eq:SFF_T_def}
\end{align}
At infinite temperature, the SFF reflects the energy correlations across all possible length scales  and is considered to be an efficient probe of quantum chaos~\cite{Dag2023, Das2025, Fritzsch2025, Kalsi2025, Roy2025, VallejoFabila2024, VallejoFabila2025, Das2026, Das2019, Das2025b}. The SFF can also be understood as the survival probability (i.e.~the probability to detect the initial state at a later time) of a specific initial state called the coherent Gibbs state (CGS), $\ket{\Psi_T}$~\cite{DelCampo2017, DelCampo2018, Xu2021}
\begin{align}
	\sff{t; T} = \abs{\mean{\Psi_T | \Psi_T(t)}}^2,\: \ket{\Psi_T} \equiv \sum_{n = 1}^{N} \sqrt{p_n(T)} \ket{\Phi_n}
	\label{eq:CGS_def}
\end{align}
where $\ket{\Phi_n}$ is the eigenstate at energy $E_n$ and $\ket{\Psi_T(t)}$ is the time evolved state. In the \phaseA~phase ($T > \Tq$), the Gibbs probability amplitude $\sqrt{p_n(T)} \to N^{-\frac{1}{2}}$ and the $\sff{t; T}$ reduces to the infinite temperature SFF. Contrarily in the \phaseF~phase ($T < \Tc$), the Gibbs state, $\ket{\Psi_T}$, converges to the ground state 
such that $\sff{t; T} \approx 1$ for all time $t$.

\begin{figure}[t]
	\centering
	\includegraphics[width=\columnwidth]{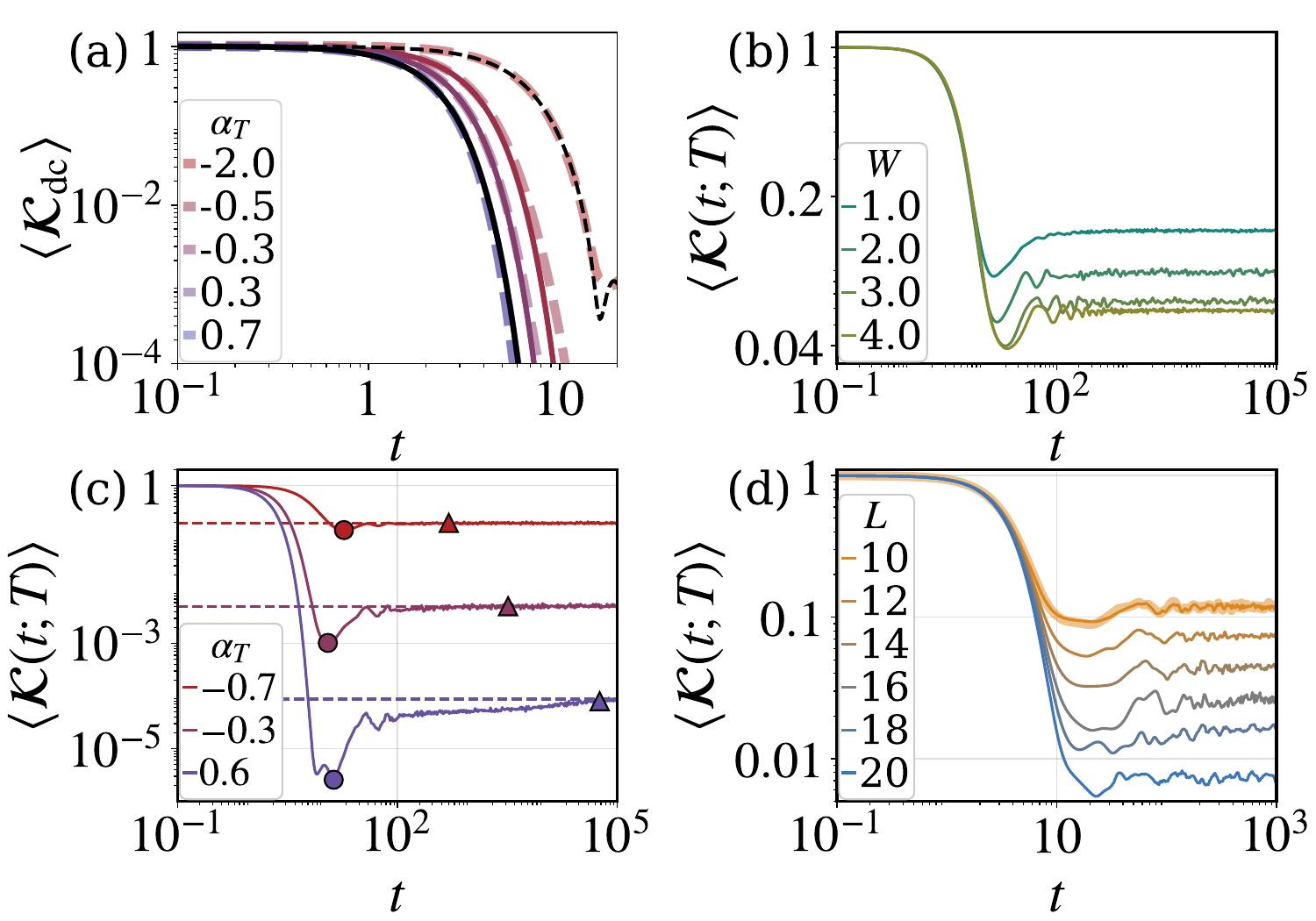}
	\caption{(a)~Disconnected part of the SFF at different temperature exponents ($\alpha_T$) for $L = 16$ and $W = 2$. The black dashed (solid) line shows the low- (high-) temperature approximate of $\mean{\sffdc{t;T}}.$
        (b)~SFF for $L=16$ and different disorder strengths at $T = 0.1$. 
		(c)~SFF for $L = 16$ and $W = 2$ and different values of $\alpha_T$. Dashed lines denote the equilibrium value, $\sffbar$. Circular (triangular) markers denote the dip (relaxation) timescales.
		(d)~SFF calculated using FTLM for $W = 5.0$, $\alpha_T = -0.5$, and different values of $L$. The thick line corresponds to the SFF computed using exact diagonalization for the smallest system size.
        }
	\label{fig:3}
\end{figure}

In general, the SFF exhibits a dip-ramp-plateau structure, with a correlation hole (a dip below the equilibrium value) in the case of a correlated energy spectrum~\cite{Leviandier1986, Pique1987, Michaille1999, Das2025, Das2025a, VallejoFabila2024, VallejoFabila2025, Das2026, Grabarits2025}. At finite temperature, there are larger Gibbs weights on the low-lying states such that the temperature acts as a filter on the energy spectrum~\cite{DelCampo2018, MatsoukasRoubeas2023}. If there exists a ME, the energy states below are localized with uncorrelated energy levels such that the SFF does not exhibit a correlation hole. Contrarily, ergodic states above the ME correspond to correlated energy levels, leading to a correlation hole in the SFF. The main result of this work is to identify the transition of relative depth of the correlation hole of the finite temperature SFF occuring at certain $\TMBL$ and relate that to the ME by exploiting the temperature-energy duality as exemplified before. 
 
Let us begin by writing the ensemble-averaged SFF  as 
\begin{align}
	\begin{split}
		\mean{\sff{t; T}} & \approx \sffbar + \mean{\sffdc{t; T}} + \mean{\sffc{t; T}}
	\end{split}
	\label{eq:SFF_T_step1}
\end{align}
where $\sffbar$ is the equilibrium value,  the disconnected [$\sffdc{t; T}$] and the connected [$\sffc{t; T}$] components are related to the Fourier transforms of local density of states (LDOS), $\rho_T(E) \equiv \sum_{n = 1}^{N} p_n \delta\del{E - E_n}$, and the two-level {correlation function}, respectively (see Appendix~\ref{sec_appnd_SFF}). A Taylor series expansion of the SFF around $t = 0$ reveals a universal quadratic decay~\cite{Wilkinson1997, Schiulaz2019, Das2023a}
\begin{align}
	\sff{t; T} = 1 - T^2\Capheat t^2 + \mathcal{O}(t^4)
	\label{eq:SFF_Zeno}
\end{align}
for very short time ($t\ll \frac{1}{T\sqrt{\Capheat}}$) where $\Capheat$ is the heat capacity [Eq.~\eqref{eq:Cv_def}]. In the \phaseA~phase, the LDOS of the CGS reduces to the DOS, while in the \phaseF~phase, the LDOS converges to the distribution of the ground state energy. Due to Gaussian shape of the DOS and ensemble equivalence, we propose the following ansatz for the LDOS in the \phaseQ~phase
\begin{align}
	\mean{\rho_T(E)} \approx \frac{1}{\sqrt{2\pi} T\sqrt{\mean{\Capheat}}} \exp\del{-\frac{(E - \mean{E}_T)^2}{2T^2\mean{\Capheat}}}.
	\label{eq:LDOS_ansatz}
\end{align}
In Fig.~\ref{fig:3}(a), we show the Fourier transform of the above ansatz along with the numerical estimate of the disconnected part of the SFF and find a nice agreement across different values of the temperature. 


The analytically continued partition function, $Z(T, it)$, involves the calculation of the trace $\mathrm{Tr}\left[e^{-(\frac{1}{T}+it)H}\right]$. To evaluate such a trace for large Hilbert-space dimensions, we employ the finite temperature Lanczos method (FTLM) by replacing the full trace with an average over \(N_v\) random vectors $\{\ket{r}\}$~\cite{Jaklic1994}. For each random vector, a Lanczos procedure with \(m\) iterations is performed, generating a tridiagonal matrix whose eigenvalues \(\{E_l^{(r)}\}\) and corresponding weights \(\{|\langle \phi_l^{(r)}|r\rangle|^2\}\) approximate the spectrum of the full Hamiltonian. The partition function is then estimated as
\begin{equation}
	Z(T, it)
	\simeq
	\frac{1}{N_v}
	\sum_{r=1}^{N_v}
	\sum_{l=1}^{m}
	|\langle \phi_l^{(r)}|r\rangle|^2
	e^{-\left(\frac{1}{T}+it\right)E_l^{(r)}}.
    \label{eq:ZT_FTLM}
\end{equation}
In practice, for each disorder realization we construct the XXX Hamiltonian in the \(S^z=0\) sector and apply a Lanczos iteration with full reorthogonalization to improve the numerical stability. The resulting tridiagonal matrix is diagonalized using standard routines for symmetric tridiagonal matrices. The SFF is subsequently obtained using Eq.~\eqref{eq:SFF_T_def} and averaged over disorder realizations. 

In Fig.~\ref{fig:3}(b), we show the variation of the SFF with disorder strength for a fixed temperature $(T = 0.1)$ whereas Fig.~\ref{fig:3}(c) shows the SFF in the ergodic phase of the Heisenberg model ($W = 2$) for different temperatures. In Fig.~\ref{fig:3}(d), we compare the SFF obtained using the FTLM with the data from exact diagonalization for the smallest $L$, while also presenting the FTLM results for larger system sizes. Note that for the exact diagonalization of an $N \times N$ matrix, typically $\mathcal{O}(N^3)$ number of operations and $\mathcal{O}(N^2)$ bytes of memory, are required which restricts the maximum attainable system size to be $L = 18$ in a typical present-day computer. In contrast, the number of operations required for the FTLM is $\mathcal{O}(N N_v m)$ where $N_v, m \ll N$, which allows us to reach up to $L = 20$, as shown in Fig.~\ref{fig:3}(d). 

\begin{figure}[t]
	\centering
	\includegraphics[width=\columnwidth]{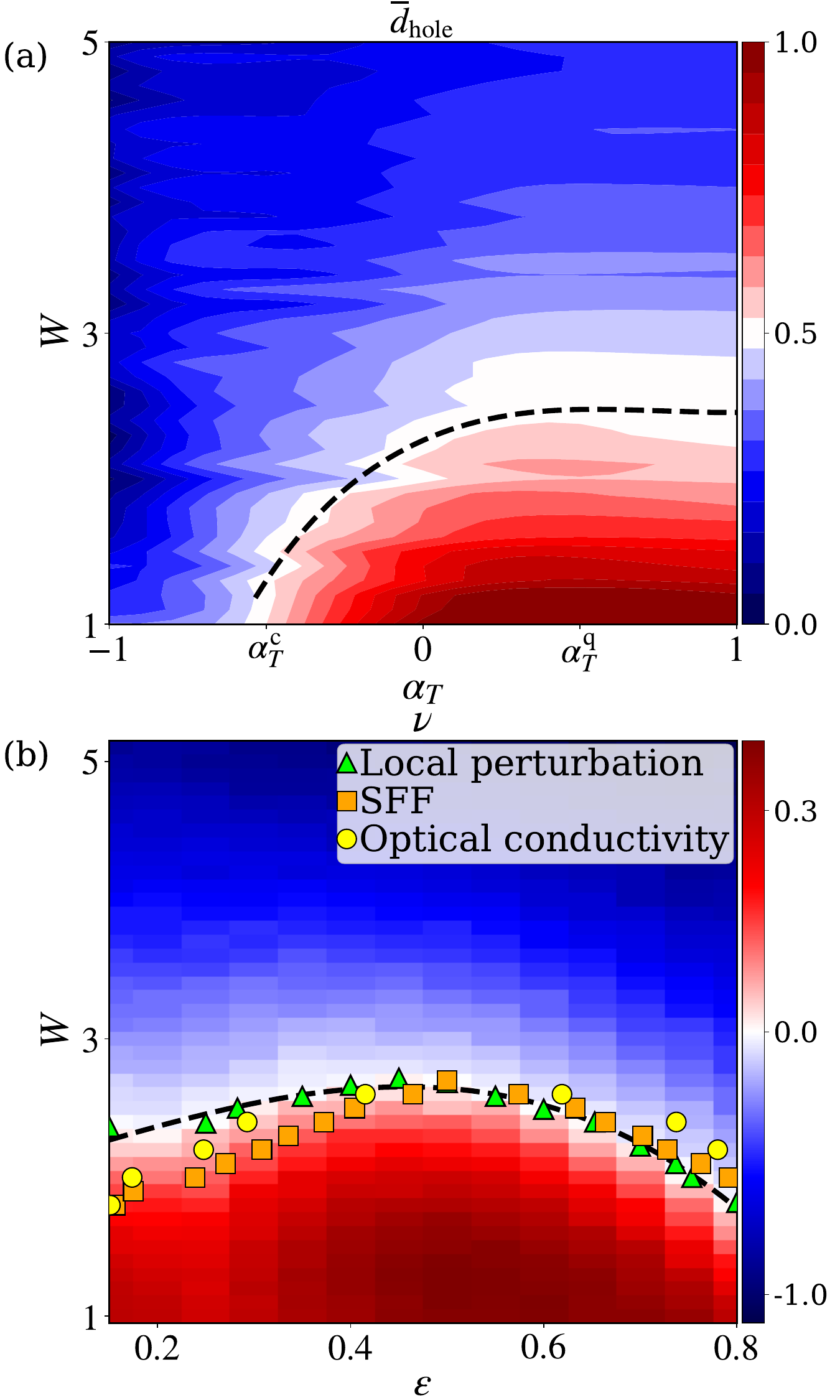}
	\caption{(a)~Canonical phase diagram: colormap of normalized $d_\mathrm{hole}$ in SFF for $L = 16$. The black dashed line corresponds to the critical temperature $\TMBL$.
		(b)~Microcanonical phase diagram: colormap of conductance exponent ($\nu$). The mobility edges obtained from the conductance exponent, the optical conductivity, and the critical energy density $\epsilon_\mathrm{MBL}$ are indicated by different markers. The quantity $\epsilon_\mathrm{MBL}$, shown for $L=16$, is dual to $\TMBL$ in panel (a) through Eq.~\eqref{eq:X_ME_MBL}. The black dashed line serves as a guide to the eye.
	}
	\label{fig:4}
\end{figure}

Now we discuss how to identify the ME using the finite temperature SFF. The SFF exhibits the minimum (dip) at the time $\tdip$ which we find to be independent of the system size and temperature. To quantify the correlations in the temperature-weighted energy spectrum, we define the relative depth of the correlation hole~\cite{Schiulaz2019, Roy2025}
\begin{align}
	d_\mathrm{hole}(W,T) = 1 - \frac{\ln\mean{\sff{\tdip}}} {\ln\sffbar}.
\end{align}
Since $T$ acts like a filter on the energy spectrum by putting larger Gibbs weights on the low excitations, $d_\mathrm{hole}$ has nontrivial dependence on both temperature and disorder strength. In Fig.~\ref{fig:4}(a) we show the normalized relative depth, 
\begin{align}
\bar{d}_\mathrm{hole} = \frac{d_\mathrm{hole}(W,T)}{d_\mathrm{hole}(W = 1,T\rightarrow \infty)} \in [0,1]
\end{align}
and define a critical line using the relation of Eq.~\eqref{eq:aT_def} as $\TMBL \equiv (\ln N)^{\alpha_T^\mathrm{MBL}}$, which corresponds to a value of 1/2. Then, in the microcanonical phase diagram, we identify the ME as the energy density
\begin{align}
	\epsilon_\mathrm{MBL} = \mean{\frac{\mean{E}_{T = T_\mathrm{MBL}} - E_1}{ E_{\mathrm{N}} - E_1}},
	\label{eq:X_ME_MBL}
\end{align}
where $E_{\mathrm{N}}$ ($E_1$) is the energy of the antiground (ground) state and the thermal mean energy is computed from Eq.~\eqref{eq:heat_cap_annealed}. 
In Fig.~\ref{fig:4}(b), we show the microcanonical phase diagram of the Heisenberg model showing $\epsilon_\mathrm{MBL}$ ($\square$ markers). In order to verify that the critical energy obtained above is indeed the ME, we compute a few physical quantities as elucidated below. 

Firstly, we perturb the Heisenberg model locally as $\hat{H} \to \hat{H} + \hat{\sigma}_{\frac{L}{2}}^z$ and compute~\cite{Serbyn2015}
\begin{align}
	\mathcal{G} = \ln \frac{|V_{n,n+1} |}{E_{n+1}^{\prime} - E_n^{\prime}}
	\label{eq:G_Abanin}
\end{align}
where $E_n^{\prime} = E_n + V_{n,n}$, $\hat{V} =  \hat{\sigma}_{\frac{L}{2}}^z$ is the local perturbation and $V_{mn} = \bra{\Phi_m} \hat{V} \ket{\Phi_n}$. The quantity $\mathcal{G}$ is analogous to the Thouless conductance in single-particle systems and scales as $\mean{\mathcal{G}} \propto \nu L$ such that the conductance exponent $\nu > 0$ ($\nu < 0$) corresponds to the ergodic (MBL) phase [see Fig.~\ref{fig:5}(a)]. Thus, the ME can be identified as the energy density at which $\langle \mathcal{G} \rangle$ becomes independent of system size $(\nu = 0)$ as shown via the $\triangle$ markers in Fig.~\ref{fig:4}(b) where the colormap shows the numerical estimates of $\nu$. In the same figure, we also show the ME estimates from the finite temperature SFF [$\epsilon_\mathrm{MBL}$ obtained from Eq.~\eqref{eq:X_ME_MBL}] and find a nice agreement.

Secondly, let us determine the optical conductivity 
\begin{align}
    \sigma(\omega, T) \equiv \frac{1 - e^{-\frac{\omega}{T}}}{\omega} \int_{0}^{t_\mathrm{max}} dt e^{i \omega t} C(t)
    \label{eq:optical_conductivity_def}
\end{align}
where $C(t)$ is the autocorrelation of the spin current, $\hat{j} = \frac{1}{4}\sum_{i} \left(\hat{\sigma}^x_{i+1} \hat{\sigma}^y_{i} - \hat{\sigma}^y_{i+1} \hat{\sigma}^x_i \right)$.
By extending the low-frequency behavior $[\sigma(\omega) = \sigma_\mathrm{dc} + |\omega|^{\eta}]$ of the optical conductivity at infinite temperature~\cite{Karahalios2009, Steinigeweg2016} to finite temperatures, we extract $\TMBL$ as the point where $\eta$ crosses $1$ [see Fig.~\ref{fig:5}(b)]. The mobility edge identified using this method is shown via $\circ$ markers in Fig.~\ref{fig:4}(b). 
The agreement of the optical conductivity measurements with the $\epsilon_\mathrm{MBL}$ (dual to $\TMBL$) once again confirms the robustness of the ME estimates from the finite temperature SFF.

\section{Discussion}\label{sec_Discussion}

In this work, we propose a method to identify the ME of a generic quantum system using finite temperature SFF and demonstrate the same in case of the 1D disordered Heisenberg spin-$\frac{1}{2}$ model. First, we identify the canonical phase diagram of the Heisenberg model using non-analyticity of the heat capacity and the system size scaling of the Gibbs weight of the ground state. We posit the existence of three thermodynamic phases in any disordered system: \phaseF~phase ($T < \Tc$) where the ground state is the only allowed configuration and shows ensemble inequivalence; \phaseQ~phase ($\Tc < T < \Tq$) where ensemble equivalence is present but equipartition is absent; and \phaseA~phase ($T> \Tq$) where ensemble equivalence holds and equipartition ensures that the equilibrium value of any local observable is given by the infinite temperature average for weak disorder strength. Such distinct thermodynamic phases get dynamically manifested in the finite temperature SFF, which does not decay in the \phaseF~phase and saturates to its infinite temperature limit in the \phaseA~phase such that the CGS spreads homogeneously over all the energy states. By looking at the relative depth of the SFF as a function of temperature, we identify the critical temperature $\TMBL$ which is dual to the ME, as verified by the spread of local perturbations and optical conducivity. Thus, our work presents a way to identify the critical energy separating the localized and extended states in the energy spectrum of the disordered systems. Our method is independent of exact diagonalization of the governing Hamiltonian, and thus provides an alternative route to identify dynamical phase transitions in many-body systems.

\begin{acknowledgements}
	We acknowledge the support from Kepler Computing facility, maintained by the Department of Physical Sciences, IISER Kolkata, for various computational needs. A.~K.~D. acknowledges support from the Leverhulme Trust Research Project Grant RPG-2025-063.
\end{acknowledgements}

\renewcommand\thefigure{\thesection.\Roman{figure}}
\setcounter{figure}{0}
\renewcommand\thetable{\thesection.\Roman{table}}
\setcounter{table}{0}
\renewcommand\theequation{\Roman{equation}}
\setcounter{equation}{0}

\appendix
\section{Energy moments}\label{sec_appnd_E_moments}
The mean energy of the Hamiltonian of the XXX model can be estimated using the relation 
\begin{align}
	\overline{E} = \frac{1}{N}\mathrm{Tr}[H],
\end{align}
where the diagonal terms only contribute and $N = \binom{L}{\frac{L}{2}}$ is the dimension of the Hilbert space. 
The diagonal contributions only come from the $z$ component of the Pauli operators
\begin{align}
	\begin{split}
		\overline{E} &= \frac{1}{4 N} \sum_{i = 1}^N \sum_{j = 1}^{L - 1} \left(1 + 4k_j(i)k_{j + 1}(i) -2k_j(i) - 2k_{j + 1}(i) \right)\\
		&+ \frac{1}{2N}\sum_{i = 1}^N \sum_{j = 1}^{L} h_j \left(1 - 2k_j(i)\right)
	\end{split}
	\label{eq:XXX_E_bar}
\end{align}
where $k_j(i)$ is the bit value $(\in \{0,1\})$ for the $j$-th site in the basis state $\ket{i}$. As each site is occupied in exactly half of the basis states, for all values of $j$ we get
\begin{align}
	\begin{split}
		\frac{1}{N} \sum_{i=1}^N k_j(i) &= \frac{1}{2}, \\
		\frac{1}{N} \sum_{i=1}^N k_j(i)\,k_{j+1}(i) &= \frac{L-2}{4(L-1)}
	\end{split}
\end{align}
and Eq.~\eqref{eq:XXX_E_bar} reduces to 
\begin{align}
	\frac{1}{4}\sum_{j = 1}^{L - 1} \left( 1 + \frac{L - 2}{L - 1} - 2 \right) = -\frac{1}{4}
\end{align}

\begin{figure}[t]
	\centering
	\includegraphics[width=\linewidth]{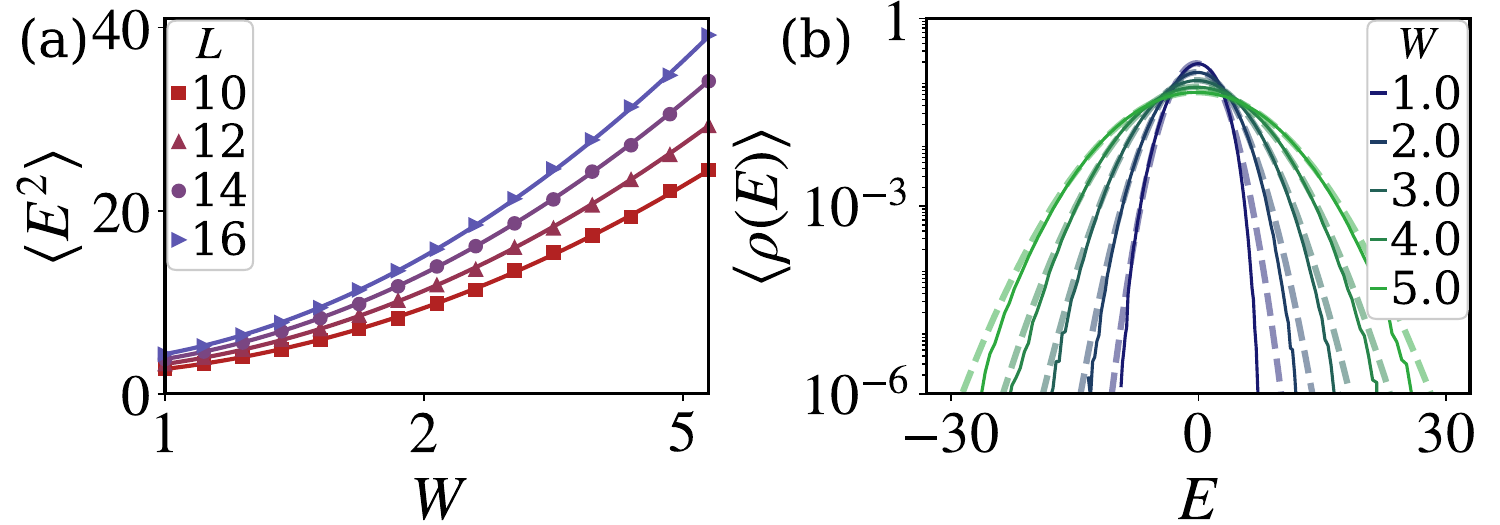}
	\caption{(a)~Second energy moment as a function of the disorder strength, $W$ for different values of $L$, the number of lattice sites. Markers denote the numerically obtained data while the solid lines denote the analytical expression from Eq.~\eqref{eq:E2_XXX}.
		(b)~DOS for $L = 16$ and different values of $W$. Solid lines denote the numerically obtained histogram and the dashed lines denote the Gaussian approximation with mean and variance determined by Eqs.~\eqref{eq:E1_XXX} and \eqref{eq:E2_XXX}.}
	\label{fig:DOS_XXX}
\end{figure}

Thus, the first moment of energy for the spin-$\frac{1}{2}$ XXX model is 
\begin{align}
	\mean{E} = -\frac{1}{4}
	\label{eq:E1_XXX}
\end{align}
irrespective of the number of spins and the disorder strength.

To calculate the second energy moment, $\overline{E^2} = \frac{1}{N} Tr[H^2]$, we denote the first and second terms of the Hamiltonian as $\hat{H}_1$ and $\hat{H}_2$ respectively. Then, the square of the matrix corresponding to the Hamiltonian becomes
\begin{align}
	H^2 = H_1^2 + H_2^2 + H_1 H_2 + H_2 H_1
\end{align}
and the diagonal contribution in both of the cross terms contain the factor coming from $H_1$ as $\sum_{j = 1}^L h_j (1 - 2k_j(i))$ which leads to
\begin{align}
	\frac{1}{N} \sum_{i = 1}^N H_1 H_2 = \frac{1}{N} \sum_{i = 1}^N H_2 H_1 = 0.
\end{align}
The remaining two terms yield
\begin{widetext}
\begin{align}
	\begin{split}
		H_1^2 + H_2^2 &= \frac{1}{16} \left(\sum_{j = 1}^{L - 1}\left(\vec{\sigma}_j\cdot \vec{\sigma}_{j+1}\right)^2 + \sum_{j \neq l}^{L - 1} \left(\vec{\sigma}_j\cdot \vec{\sigma}_{j+1}\right) \left(\vec{\sigma}_l\cdot \vec{\sigma}_{l+1}\right) \right) + \frac{1}{4}\left(\sum_{j = 1}^L h_j \sigma^z_j \right)^2 \\
		&= \frac{1}{16} \sum_{j = 1}^{L - 1} \left[ 
		\left(\sigma_{j}^x\sigma_{j+1}^x \right)^2 
		+ \left(\sigma_{j}^y\sigma_{j+1}^y \right)^2 
		+ \left(\sigma_{j}^z\sigma_{j+1}^z \right)^2 
		+ \left\{\sigma_{j}^x\sigma_{j+1}^x,~\sigma_{j}^y\sigma_{j+1}^y \right\} 
		+ \left\{\sigma_{j}^y\sigma_{j+1}^y,~\sigma_{j}^z\sigma_{j+1}^z \right\} 
		+ \left\{\sigma_{j}^z\sigma_{j+1}^z,~\sigma_{j}^x\sigma_{j+1}^x \right\} 
		\right] \\
		&+ \frac{1}{16}\sum_{j \neq l}^{L - 1}\left[\left( \sigma_{j}^x\sigma_{j+1}^x +\sigma_{j}^y\sigma_{j+1}^y + \sigma_{j}^z\sigma_{j+1}^z\right)\left( \sigma_{l}^x\sigma_{l+1}^x +\sigma_{l}^y\sigma_{l+1}^y + \sigma_{l}^z\sigma_{l+1}^z\right)\right] + \frac{1}{4}\left[\sum_{j = 1}^L \left(h_j \sigma^z_j\right)^2 + \sum_{j \neq l}^{L}\left(h_j h_l \sigma^z_j \sigma^z_l\right)\right] \\
		&= \frac{1}{16} \left[\sum_{j = 1}^{L - 1} \left(3 \mathbb{I}
		- \frac{1}{2}\left( \sigma_{j}^x\sigma_{j+1}^x + \sigma_{j}^y\sigma_{j+1}^y + \sigma_{j}^z\sigma_{j+1}^z\right) 
		\right)\right] \\
		&+ \frac{1}{16}\sum_{j \neq l}^{L - 1}\left[\left( \sigma_{j}^x\sigma_{j+1}^x +\sigma_{j}^y\sigma_{j+1}^y + \sigma_{j}^z\sigma_{j+1}^z\right)\left( \sigma_{l}^x\sigma_{l+1}^x +\sigma_{l}^y\sigma_{l+1}^y + \sigma_{l}^z\sigma_{l+1}^z\right)\right] + \frac{1}{4}\sum_{j = 1}^{L} h_j^2 \mathbb{I} + \frac{1}{4}\sum_{j \neq l}^{L}\left(h_j h_l \sigma^z_j \sigma^z_l\right)
	\end{split}
\end{align}
\end{widetext}
In the above expression, considering only the diagonal terms, the mean of $E^2$ becomes
\begin{align}
	\begin{split}
		\overline{E^2} &= \frac{3}{16}(L-1) + \frac{1}{8} + \frac{1}{16} + \frac{1}{4}\sum_{j = 1}^{L} h_j^2 = \frac{3}{16}L + \frac{1}{4}\sum_{j = 1}^{L} h_j^2. 
	\end{split}
\end{align}
Thus, we obtain the second moment of energy
\begin{align}
	\mean{E^2} = \frac{3}{16}L + \frac{W^2}{12}L.
	\label{eq:E2_XXX}
\end{align}
In Fig.~\ref{fig:DOS_XXX}(a), we show the second moment of the energy as a function of disorder strength along with the analytical expression in Eq.~\eqref{eq:E2_XXX} for different system sizes and find excellent agreement.

To obtain a microcanonical description of a system, we need to understand the corresponding DOS. In Fig.~\ref{fig:DOS_XXX}(b), we compare the DOS of the XXX model for various values of $W$ with Gaussian distributions having the same mean and variance as obtained analytically in Eqs.~\eqref{eq:E1_XXX} and \eqref{eq:E2_XXX}. For finite system sizes, the DOS is negatively skewed and becomes more symmetric as we increase the disorder strength. Nevertheless, we can assume the DOS of the XXX model to be Gaussian and analytically estimate the critical temperatures, as shown in the main text.

\begin{figure}
	\centering
	\includegraphics[width=0.75\linewidth]{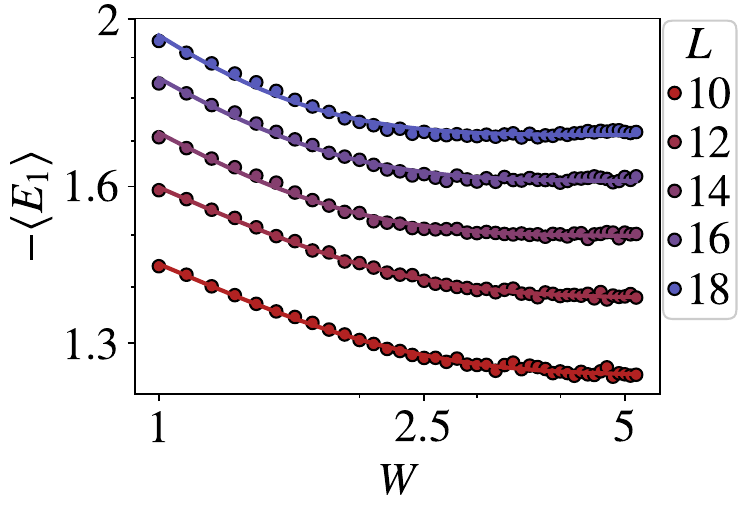}
	\caption{Exponential fit of the mean of the ground state energy. Circles show the numerical values of $-\mean{E_1}$ and solid lines show the fitted exponential form as given in Eq.~\ref{eq:param_vals}.}
	\label{fig:E1_fit}
\end{figure}

Apart from the DOS, we also need to estimate the scaling of the mean of the ground state energy, $\mean{E_1}$, as a function of the system size, $L$, and the disorder strength, $W$. As shown in Fig.~\ref{fig:E1_fit}, the mean can be approximated with an exponentially decaying function: $\mean{E_1} \approx -a(L) e^{-b(L) W} - c(L)$, where the fit parameters $a, b$ and $c$ are
\begin{align}
	\begin{split}
		a(L) &\approx 0.015\,L^{1.61}, \\
		b(L) &\approx 0.1\,L + 0.2, \\
		c(L) &\approx 0.36\,L^{0.5}.
	\end{split}
	\label{eq:param_vals}
\end{align}

\section{Finite temperature spectral form factor}\label{sec_appnd_SFF}
The finite temperature SFF can be given in terms of the analytical continuation of the partition function as 
\begin{align}
	\begin{split}
		\sff{t; T} &\equiv \frac{\abs{\nrm(T,it)}^2}{\nrm(T)^2} = \frac{ \sum_{m, n}^{N} e^{i(E_m-E_n)t - \frac{E_m+E_n}{T}} }{\nrm(T)^2},
	\end{split}
	\label{eq_Apnd:SFF_T_def}
\end{align}
where $t$ is the time and $Z(T) \equiv Z(T,0)$. We can perform an ensemble average of Eq.~\eqref{eq:SFF_T_def} and the ensemble-averaged SFF is given by
\begin{align}
	\begin{split}
		\mean{\sff{t; T}} &= \sffbar + \int d\vec{E} \prob{\vec{E}} \sum_{m \neq n} \frac{\exp\del{-\frac{E_{mn}}{T} - i\Omega_{mn} t}}{\nrm(T)^2}
	\end{split}
	\label{eq_Apnd:SFF_T_step1}
\end{align}
where $E_{mn} \equiv E_m + E_n$, $\Omega_{mn} \equiv E_m - E_n$ and $\sffbar$ is the equilibrium value of the SFF at a temperature $T$ defined as 
\begin{align}
	\begin{split}
		\sffbar &\equiv \mean{\frac{\nrm(T/2)}{\nrm(T)^2}}
	\end{split}
	\label{eq:SFF_T_bar}
\end{align}
and $\prob{\vec{E}}$ is the joint density of the energy levels. At high temperatures where energy ordering is not important we can use annealed approximation and decompose the ensemble-averaged SFF into a connected and a disconnected part as
\begin{align}
	\begin{split}
		\mean{\sff{t; T}} & \approx \sffbar + \mean{\sffdc{t; T}} + \mean{\sffc{t; T}}.
	\end{split}
	\label{eq_Apnd:SFF_T_step2}
\end{align}
The first and the second terms are respectively
\begin{widetext}
\begin{align}
	\begin{split}
		\mean{\sffdc{t; T}} &\equiv \frac{N^2}{\mean{\nrm(T)}^2} \int dE_m dE_n \mean{\rho(E_m)} \mean{\rho(E_n)} \exp\del{-\frac{E_{mn}}{T} - i\Omega_{mn} t},\\
		\mean{\sffc{t; T}} &\equiv -\frac{1}{\mean{\nrm(T)}^2} \int dE_m dE_n T_2(E_m, E_n) \prob{E_{mn}} \exp\del{-\frac{E_{mn}}{T} - i\Omega_{mn} t}, 
	\end{split}
\end{align}
\end{widetext}
where $T_2(E_m, E_n)$ is the two-level cluster function related to the two-point energy correlation, $\rho^{(2)}(E_m, E_n)$ as
\begin{align}
	\mean{\rho^{(2)}(E_m, E_n)} = N^2 \mean{\rho(E_m)} \mean{\rho(E_n)} - T_2(E_m, E_n).
	\label{eq:rho_2_def}
\end{align}
\section{Spread of local perturbation}
\begin{figure}[t]
	\centering
	\includegraphics[width=\columnwidth]{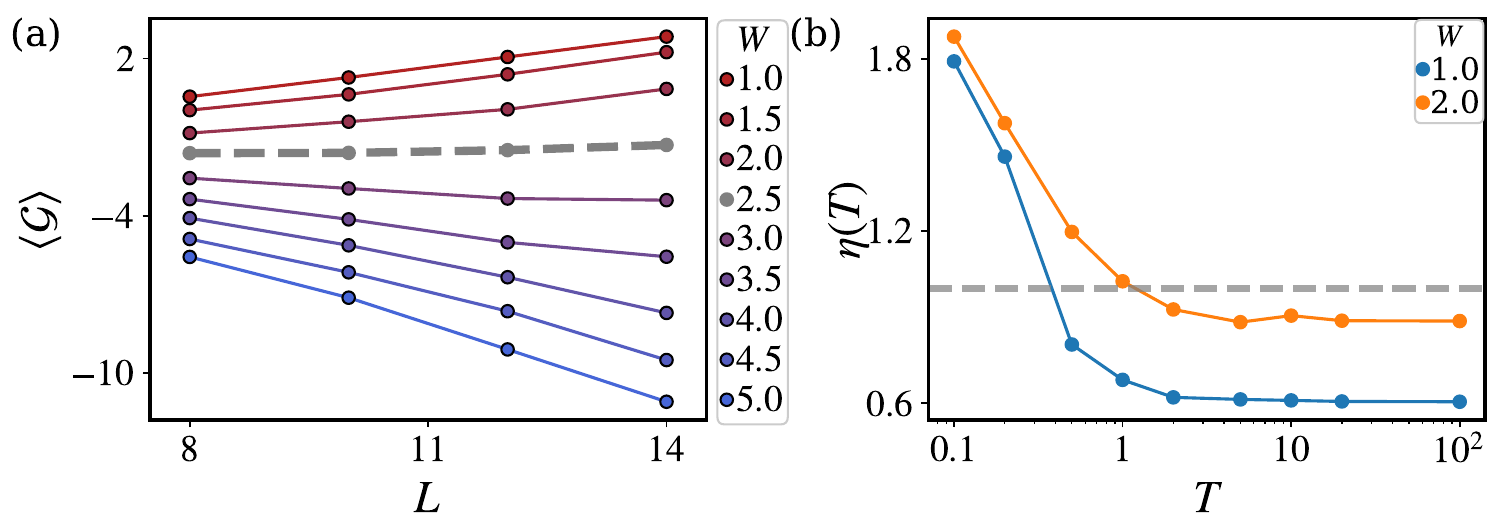}
	\caption{(a)~Variation of the quantity $\mean{\mathcal{G}}$ with $L$ for different disorder strengths at energy density $\epsilon = 0.5$. The gray dashed line denotes the critical value of the disorder strength $W_c$.  
		(b)~Temperature dependence of the exponent $\eta$ for two different disorder strengths. The horizontal dashed line marks the value $\eta = 1$.
	}
	\label{fig:5}
\end{figure}

In Ref.~\cite{Serbyn2015}, the response of quantum many-body systems to a local perturbation and its use as a probe to ergodicity is discussed. To analyze the modification of the eigenstates under the influence of the local perturbation, a parameter $\mathcal{G}$, similar to Thouless conductance in the single-particle localization, is introduced.
The parameter is defined in Eq.~\eqref{eq:G_Abanin}. The ergodic and MBL phases can be distinguished by examining the mean of the distribution of $\mathcal{G}$ as a function of the number of spins $L$. The local operator in the MBL phase couples the eigenstates exponentially weakly in $L$, and as a result, in the MBL phase, $\mean{\mathcal{G}} \propto -L$.  However, the ETH implies that the energy levels in the ergodic phase are significantly mixed by a local disturbance, requiring $\mean{\mathcal{G}} \propto L$. Thus, the many-body mobility edge can be resolved by identifying the critical disorder strength $W_c$ where $\mean{\mathcal{G}}$ becomes independent of $L$, for a given energy density $\epsilon$ [defined as $\epsilon = (E - E_1) / (E_\mathrm{N} - E_1)$]. In Fig.~\ref{fig:5}(a) we show the variation of $\mean{\mathcal{G}}$ with $L$ for different values of disorder strength and identify $W_c$ from it at a fixed value of $\epsilon$. For various values of $\epsilon$ we linearly fit the data of $\mean{\mathcal{G}}$ vs $L$ as $\mean{\mathcal{G}} = \nu L + \mathcal{G}_c$ and obtain the phase diagram of $\nu$ in the $\epsilon-W$ plane.

\section{Optical conductivity}
In systems that host a many-body mobility edge, finite temperature transport is governed by thermally activated contributions from delocalized eigenstates lying above the mobility edge. In this sense, temperature acts as an effective filter that selectively weights extended states. A particularly useful probe of localization properties at finite temperatures is the frequency- and temperature-dependent optical conductivity, which within linear response theory is defined as in Eq.~\eqref{eq:optical_conductivity_def}. \( \omega \) denotes the angular frequency of the external probe field coupled to the current operator, and \( t_{\mathrm{max}} \) is a cutoff time chosen to be much larger than the relaxation time of the current autocorrelation function \( C(t) \) \cite{Steinigeweg2016,Steinigeweg2014a}.
The autocorrelation function is defined as
\begin{align}
	C(t) = \frac{\langle \hat{j}(t) \hat{j} \rangle}{L}
	= \frac{\mathrm{Tr}\!\left[e^{-H/T} \hat{j}(t) \hat{j} \right]}{L\,\mathrm{Tr}\!\left[e^{-H/T}\right]},
\end{align}
where \( \hat{j} = \frac{1}{4}\sum_{i} \left(\hat{\sigma}^x_{i+1} \hat{\sigma}^y_{i} - \hat{\sigma}^y_{i+1} \hat{\sigma}^x_i \right) \) is the spin current operator and \( L \) is the system size.
The low-frequency behavior of the optical conductivity provides a clear distinction between diffusive, subdiffusive, and localized transport regimes, and is therefore widely used as a diagnostic of many-body localization and its dynamical signatures \cite{Gopalakrishnan2015, Agarwal2015}. In the infinite temperature limit, the conductivity at low frequencies follows the form
\[
\sigma(\omega) = \sigma_{\mathrm{dc}} + |\omega|^{\eta},
\]
where \( \sigma_{\mathrm{dc}} \) is the dc conductivity and the exponent \( \eta \) varies continuously across the MBL phase \cite{Karahalios2009,Steinigeweg2016}. At the MBL transition, \( \eta = 1 \), while deep in the localized phase it approaches \( \eta \to 2 \) \cite{Gopalakrishnan2015}. We use this phenomenology in the finite temperature regime by allowing the exponent \( \eta \) to depend explicitly on temperature. By analyzing the temperature dependence \( \eta(T) \), we identify the characteristic temperature \( T_{\mathrm{MBL}} \) associated with the MBL transition for different disorder strengths.

Several numerical techniques have been developed to study transport properties in interacting quantum systems, including ED \cite{narozhny1998,heidrich2003,herbrych2011,steinigeweg2009,steinigeweg2011}, the FTLM \cite{Jaklic1994,Jaklic1994a,Jaklic2000}, the low-temperature Lanczos method \cite{aichhorn2003}, and dynamical quantum typicality (DQT) \cite{Steinigeweg2014b,Steinigeweg2016}. In this work, we employ the DQT approach, which exploits the fact that a single randomly chosen pure state can accurately reproduce ensemble-averaged properties in large Hilbert spaces \cite{elsayed2013,Steinigeweg2014b}.

Within the DQT framework, the current autocorrelation function is evaluated as
\begin{align}
	C(t) = \mathrm{Re}\,
	\frac{\langle \Phi_T(t) | \hat{j} | \phi_T(t) \rangle}
	{L \langle \Phi_T(0) | \Phi_T(0) \rangle}
	+ \delta,
	\label{eq:Ct_dqt}
\end{align}
where \( |\Phi_T(t)\rangle = e^{-iHt - H/(2T)}|\psi\rangle \) and
\( |\phi_T(t)\rangle = e^{-iHt} \hat{j} e^{-H/(2T)}|\psi\rangle \).
Here, \( |\psi\rangle \) is a random pure state, and the statistical error term is exponentially suppressed with increasing system size at any temperature. Time evolution of the states is carried out using a fourth-order Runge-Kutta scheme.
In Fig.~\ref{fig:5}(b), we show the temperature dependence of the exponent \( \eta \). We identify \( T_{\mathrm{MBL}} \) as the temperature at which \( \eta(T) \) crosses unity. To determine the corresponding mobility edge we use the relation given in Eq.~\eqref{eq:X_ME_MBL}. This allows us to determine the mobility edge as a function of disorder strength. For energy densities satisfying \( \langle E \rangle_T > \overline{E} \), we invoke the concept of negative temperatures, which is well defined for systems with bounded spectra and has been extensively discussed in earlier studies \cite{ramsey1956,mondragon2015,abraham2017}.

\bibliography{ref_SFF_T_MBL}

\end{document}